\documentstyle[aps,epsfig,epsf]{revtex}
\def\be{\begin{equation}}
\def\ee{\end{equation}}

\def\PL{{ Phys.\ Lett.\ }}
\def\PR{{ Phys.\ Rev.\ }}

\def\PRL{{ Phys.\ Rev.\ Lett.\ }}

\newcommand{\AmS}{{\protect\the\textfont2
  A\kern-.1667em\lower.5ex\hbox{M}\kern-.125emS}}
\title{Sequential Quarkonium Suppression
}
\author{
S. Digal, P. Petreczky and H. Satz\\
Fakult\"at f\"ur Physik, Universit\"at Bielefeld,
P.O. Box 100131, D-33501 Bielefeld, Germany
}
\begin{document}
\maketitle

\begin{abstract}
We use recent lattice data on the heavy quark potential in order
to determine the dissociation temperatures of different quarkonium
states in hot strongly interacting matter. Our analysis shows in
particular that certain quarkonium states dissociate below the
deconfinement point.
\end{abstract}

\section{Introduction}
The behavior of the heavy quarkonium states in hot strongly
interacting matter was proposed as test of its confinement status, 
since a sufficiently hot deconfined medium will dissolve  
any binding between the quark-antiquark pair \cite{matsui86}.
Another possibility of dissociation of certain quarkonium states
(subthreshold states at $T=0$) is the decay into open charm (beauty)
mesons due to in-medium modification of quarkonia and heavy-light
meson masses \cite{digal01a}. 

The production of  $J/\psi$ and $\Upsilon$ mesons 
in hadronic reactions occurs in part
through production of higher excited $c \bar c$ (or $b \bar b$) states 
and their decay into quarkonia ground state. Since the lifetime of
different subthreshold quarkonium states is much larger than the
typical life-time of the medium which may be produced in nucleus-nucleus
collisions their decay occurs almost completely outside the produced
medium. This means that the produced medium can be probed not only by
the ground state quarkonium but also by different excited quarkonium states.
Since different quarkonium states have different sizes ( binding energies ), 
one expects that higher excited states will dissolve at smaller 
temperature than the smaller and more tightly bound ground states.
These facts may lead to a sequential suppression pattern in $J/\psi$ and
$\Upsilon$ yield in nucleus-nucleus collision as the function of the
energy density.

Here we will discuss heavy quarkonium dissociation below the
deconfinement point where it is due to in-medium modification of the
open charm (beauty) threshold \cite{digal01a} as well as above 
the deconfinement point due to the well known
screening phenomenon \cite{digal01b}.

\section{Quarkonium production and feed-down}

It is well known that $J/\psi$ production in hadron-hadron collision is to 
a considerable extent due to the production and subsequent decay of higher
excited $c \bar c$ states \cite{cobb78,lem82,ant92}. The feed-down from
higher excited states was systematically studied in proton-nucleon and
pion-nucleon interactions with $300 GeV$ incident proton (pion) beams 
\cite{ant92}. In these studies the cross sections for
direct production of different charmonium states (excluding feed-down)
were measured. Then making use of the known branching ratios 
$B[\chi_1(1P) \to \psi(1S)] = 0.27
\pm 0.02$, $B[\chi_2(1P) \to \psi(1S)] = 0.14 \pm .01$, and
$B[\psi(2S) \to \psi(1S)] = 0.55 \pm 0.05$, one obtains the fractional
feed-down contributions $f_i$ of the different charmonium states to the
observed $J/\psi$ production; these are shown in the second and third columns of
Tab. 1 

In the case of bottomonium the experiment provides only the inclusive
(i.e. including also the feed-down from higher states) cross section 
for different $(nS)$ states \cite{cdf}. The feed-down from $(nP)$ states is known
only for transverse momenta $p_T \ge 8 GeV/c$ \cite{cdf}. To analyze the complete
feed-down pattern, we thus have to find a way to extrapolate these data
to $p_T=0$ as well as to determine the direct cross section for
different $(nS)$ states. This can be done using the most simple
and general model for quarkonium production, the color evaporation model \cite{mangano}.
In particular this model predicts that the ratios of cross sections for production
of different quarkonium states are energy independent. This prediction
was found to be true for a considerable range of energies \cite{gavai95}. 
The ratios between the different $\chi_l(1P)$ states in this model are predicted to be
governed essentially by the orbital angular momentum degeneracy
\cite{mangano}; we thus expect for the corresponding cross-sections
\be
\chi_0(1P) : \chi_1(1P) : \chi_2(1P) = 1 : 3 : 5.
\label{2.3}
\ee
From Table 1 we have for $\pi^-N$ collisions $\chi_2(1P) / \chi_1(1P)
\simeq 1.44 \pm 0.38$ and thus reasonable agreement with the predicted
ratio 1.67. Actually, for $pN$ interactions, the experiment measures only the
combined effect of $\chi_1$ and $\chi_2$ decay (30 \% of the overall
$J/\psi$ production); the listed values in Tab.~1 are obtained by distributing this in
the ratio 3:5.

Using considerations based on color evaporation model, in particular
Eq. (1), the feed-down from higher excited $b \bar b$ states to $\Upsilon$
production can be predicted \cite{digal01b}; the feed-down fraction are summarized 
in Tab.1. Alternatively the feed-down fraction from higher excited $b \bar b$
states can be predicted using NRQCD factorization formula \cite{digal01b}.
The results of this analysis are summarized in the last column of Tab. 1

\begin{center}
\begin{tabular}{|c|c|c||c|c|c|}
\hline
      &          &              &              &          &     \\
state & $f_i(\pi^-N)$i [\%] & $f_i(p~N)$ [\%]& state & $f_i(p\bar p)$ [\%]
& $f_i^{NRQCD}(p \bar p)$ [\%] \\
&  &  & & &  \\
\hline
\hline
&  &  & & &  \\
$J/\psi(1S)$ & 57 $\pm$ 3 & 62 $\pm$ 4  & $\Upsilon(1S)$
& 52 $\pm$ 9 & 52 $\pm$ 34  \\
&  &  & & & \\
\hline
&  &  & & &  \\
$\chi_1(1P)$  & 20 $\pm$ 5  & 16 $\pm$ 4  & $\chi_b(1P)$ &
26 $\pm$ 7 & 24 $\pm$ 8  \\
&  &  & & &  \\
\hline
&  &  & & &  \\
$\chi_2(1P)$ & 15 $\pm$ 4 & 14 $\pm$ 4 & $\Upsilon(2S)$ & 10
$\pm$ 3
& 8 $\pm$ 7  \\
& &  & & & \\
\hline
& &  & & & \\
$\psi(2S)$ & 8 $\pm$ 2 & 8 $\pm$ 2 & $\chi_b(2P)$ & ~10 $\pm$ 7
&
14 $\pm$ 4 \\
& &  & & & \\
\hline
\hline
& &  & & & \\
 &  &  & $\Upsilon(3S)$
 & 2 $\pm$ 0.5  & 2 $\pm$ 2\\
& &  & & & \\
\hline
\end{tabular}
\end{center}

\par
\centerline{TAB 1: Feed-down fractions for from higher excited states to
the $J/\psi$ and $\Upsilon$ states.}
\bigskip

\section{Quarkonium dissociation below  deconfinement}
Recent lattice calculations of the heavy quark potential 
show evidence for the string breaking at finite temperature \cite{karsch01}.
On the lattice the potential is calculated from the Polyakov loop correlator,
to which it is related by 
\be
V(r,T)=-\ln< L(r) L^{\dagger}(0)>+C,
\label{polcorr}
\ee
where $L(r)$ is the Polyakov loop (see e.g. Ref. \cite{karsch01}
for definition). 
The normalization constant $C$ contains both the cut-off dependent self-energy
and the entropy contributions $-TS$ (for $T \ne 0$ $-\ln< L(r) L^{\dagger}(0)>$ is
actually the free energy of the static $Q \bar Q$ pair).
For a properly chosen
normalization constant $C$, $V(r,T)$ is the ground state energy of 
an infinitely heavy $Q \bar Q$ pair.
In absence of dynamical quarks (quenched QCD) $V(r,T)$ is linearly rising with
$r$ for large separations indicating the existence of a flux tube (string).
If dynamical quarks are present the flux tube can decay (the string can break)
by creating a pair
of light quarks $q \bar q$ from the vacuum once $V(r,T)$ is larger then twice
the binding energy of a heavy-light $Q \bar q$ ($q \bar Q$) meson
\footnote{Similar phenomenon occurs of course at $T=0$. However it is 
much more difficult to observe it on lattice (see e.g. \cite{bali00}).}.
Thus the potential at very large distances is constant $V_{\infty}(T)$ and is
equal to twice the binding energy of a heavy-light meson. 

\begin{figure}
\vspace*{-0.5cm}
\epsfxsize=8cm
\centerline{\epsffile{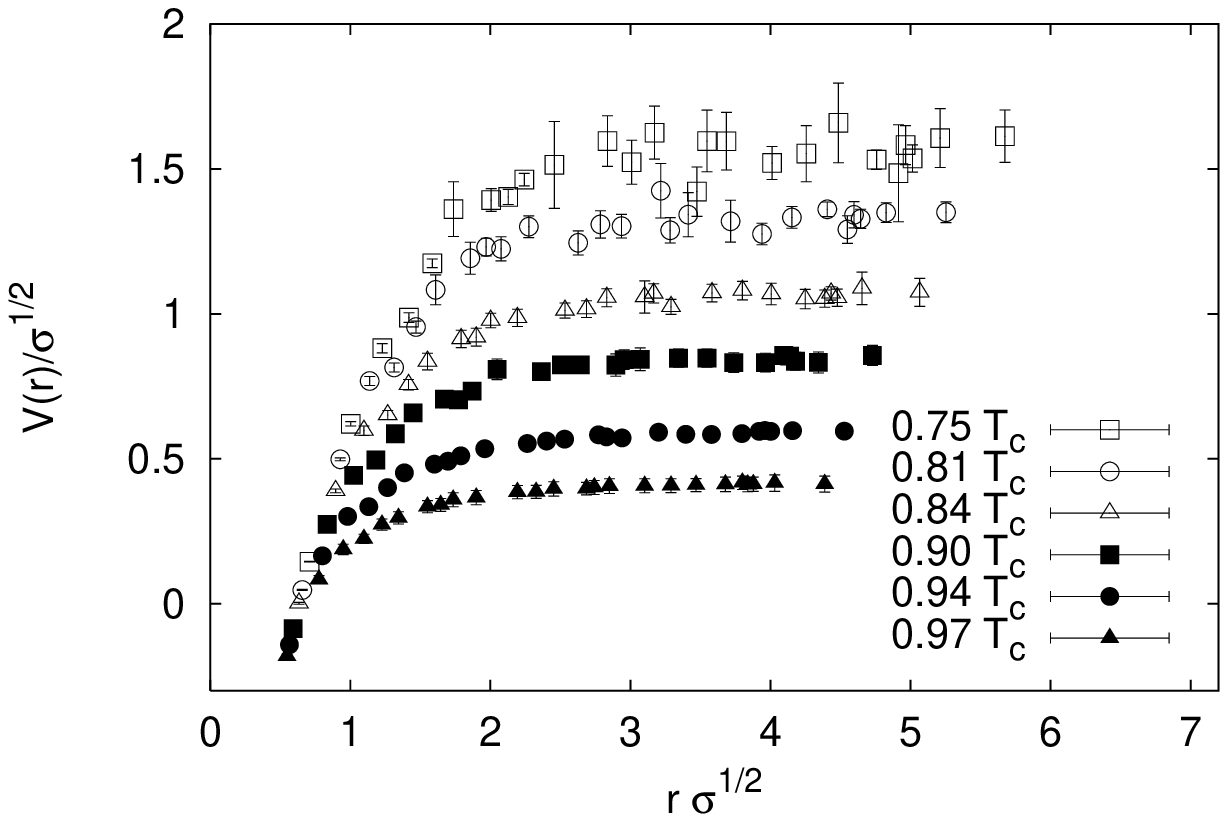} \hspace*{0.4cm} \epsfxsize=8cm \epsffile{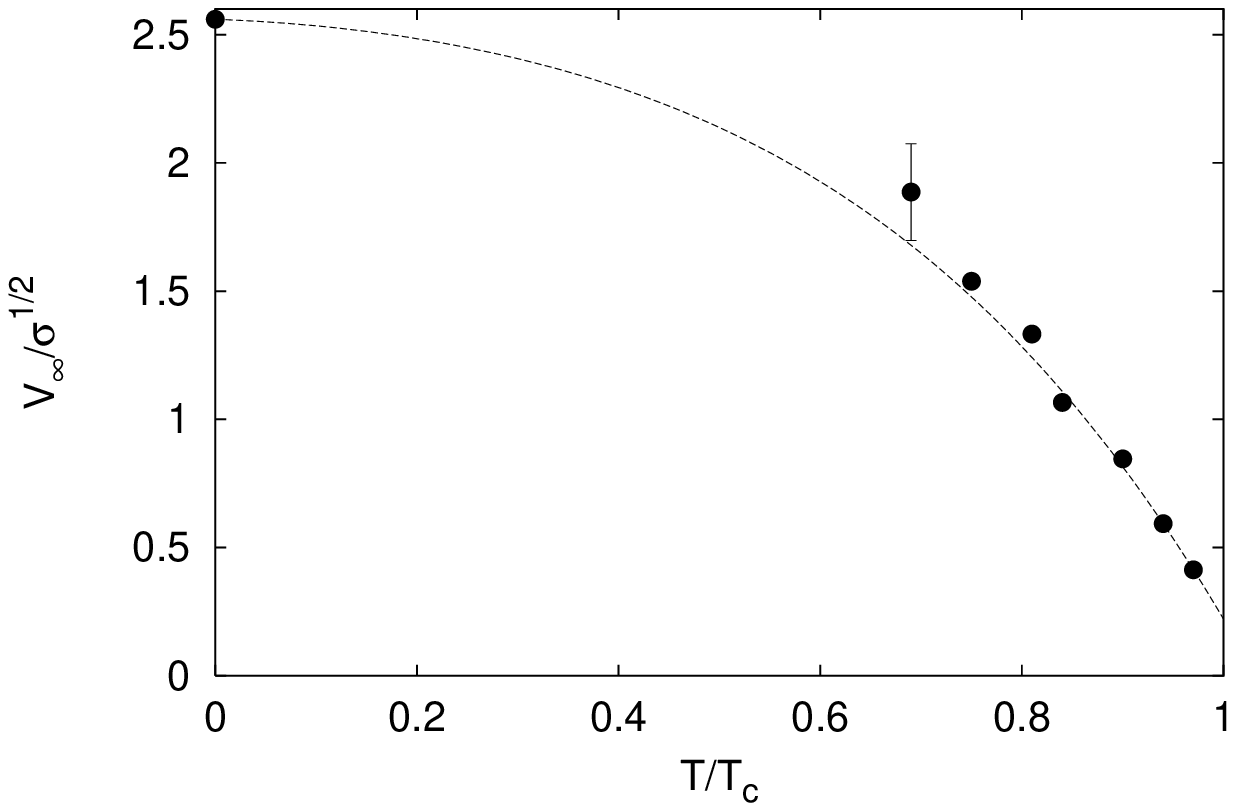}}
\vspace*{-0.3cm}
\caption{The heavy quark potential and its asymptotic value below deconfinement at
different temperatures. The line on the left figure is a fit to the data
points.}
\label{pot_below}
\end{figure}

It is expected that medium effects are not important at very
short distances. Therefore at very short distances the potential
$V(r,T)$ should be given by the Cornell potential \cite{eichten80}
\be
V(r)=-\frac{e}{r}+\sigma r 
\label{cornell}
\ee
We use this fact to determine the normalization constant $C$ and set the 
potential to be of the Cornell form at the the smallest distance $r T=0.25$ available
in lattice studies of \cite{karsch01},
with $e=0.4$ as expected for (2+1)-flavor QCD \cite{bali97}.
The resulting potential and $V_{\infty}(T)$ are shown in Fig. \ref{pot_below}.
Note the strong temperature dependence of $V_{\infty}(T)$.
Since for
sufficiently heavy quarks ($m_Q \gg \Lambda_{QCD}$) it does not matter whether the quark
is infinitely heavy or just merely heavy, the open charm (beauty) meson masses are
approximately given by $2 M_{D,B}(T)=2 m_{c,b}+V_{\infty}(T)$. 

\begin{figure}
\vspace*{-0.5cm}
\epsfxsize=8cm
\centerline{\epsffile{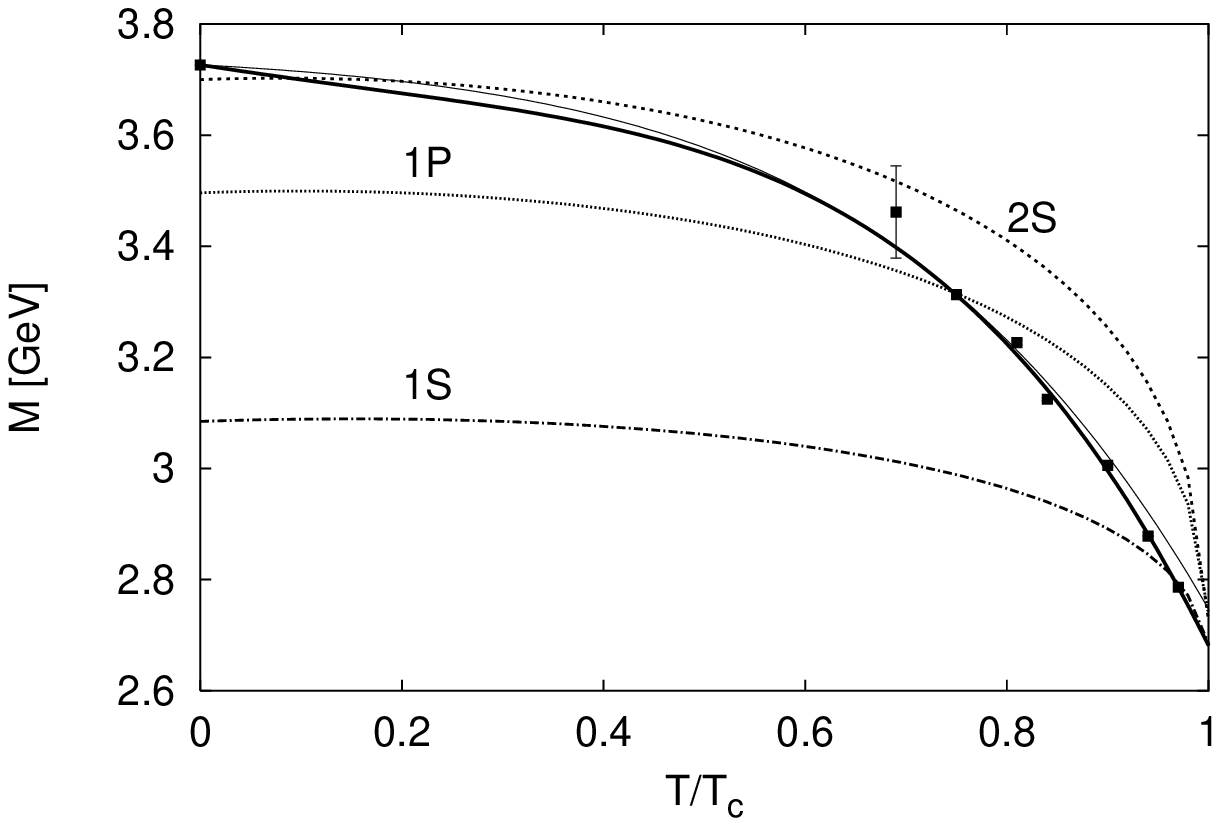} \hspace*{0.4cm}\epsfxsize=8cm \epsffile{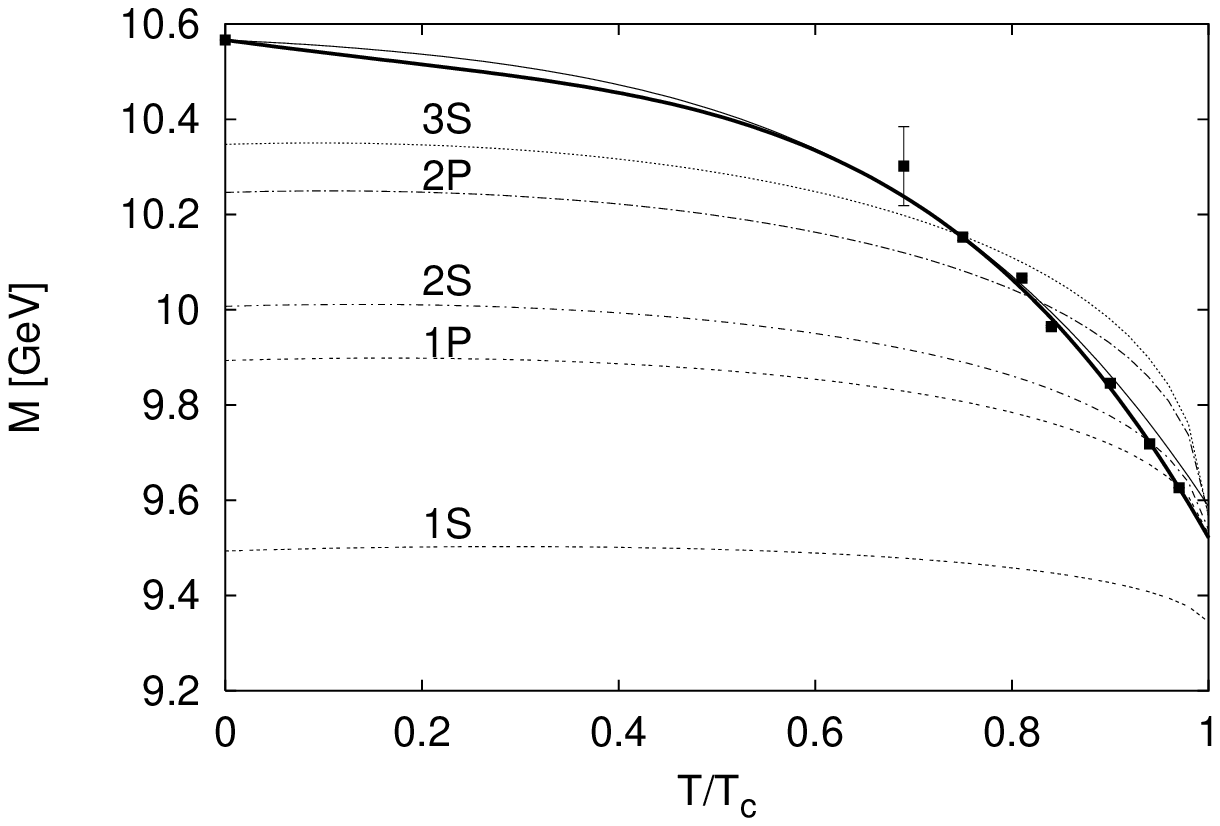}}
\vspace*{-0.3cm}
\caption{The masses of different quarkonia states and the open charm (beauty)
threshold as function of the temperature. Shown are the charmonia masses and 
open charm threshold (left) and bottomonia masses and open beauty threshold
(right) as function of the temperature. The thick solid line is the open charm
(beauty) threshold obtained from normalization at $r=1/(4T)$. The thin solid line
in the open charm (beauty) threshold obtained from normalization at $r=\sqrt{2}/(4T)$
(see text).}
\label{th}
\end{figure}

Now the temperature dependence of the different quarkonium states should
be addressed. At zero temperature the heavy quark masses permit an
application of potential theory for description quarkonium spectroscopy
(see e.g. \cite{bali00}). Furthermore it turns out that the time scale
of gluodynamics relevant for quarkonium spectroscopy is smaller
than ${(m_Q v)}^{-1}$ ($v$ being the heavy quark velocity) \cite{bali00}. 
For sufficiently
heavy quarks this time scale is much larger than the typical hadronic
time scale $\Lambda_{QCD}^{-1} \sim 1 fm$. The decay of the flux-tube
like all other hadronic decays has time scale of order $1 fm$. Therefore
in the potential theory the potential must always  be of Cornell form 
(i.e. linearly rising at large distances). These considerations have direct
phenomenological support. Namely, simple potential models with linearly 
rising potential can describe reasonably well also the quarkonium states 
above the open charm (beauty) threshold. Many of this higher excited
states have effective radius of order or even larger than $1fm$ \cite{bali97,eichten80}.
Contrary to this situation in the case of the potential becoming flat
around $1 fm$ (the expected radius of string breaking at $T=0$) the higher excited states
above the open charm (beauty) threshold simply do not exist.
Therefore we have determined the temperature dependent heavy quarkonium masses
from the Schr\"odinger equation with the temperature dependent string potential
\cite{gao}
\begin{eqnarray}
V_{string}(r,T)&=&-(e-{1\over 6} {\rm arctan}(2 r T) )~{1\over r}+
\nonumber\\
&&
(\sigma(T)-{\pi T^2\over 3}-{2 T^2\over 3} {\rm arctan}{1\over r T})~r+
{1\over 2} \ln(1+4 r^2 T^2).
\label{vTdep}
\end{eqnarray}
This form of the potential describes quite well the temperature
dependence of the heavy quark potential in quenched QCD  for appropriately
chosen $\sigma(T)$ \cite{kaczmarek00}. In order to make contact to real QCD we set $\alpha_s=0.4$
and use $T_c/\sqrt{\sigma}=0.425$ from \cite{karsch01}
for the deconfinement temperature ( by $\sigma$ we always
denote the string tension at zero temperature). 
Furthermore we use the following values of the heavy quark masses, $m_c=1.3 GeV$
and $m_b=4.72 GeV$ as well as $\sqrt{\sigma}=0.44GeV$ for the zero temperature
string tension. 
This values of parameters give a fairly good description of the observed quarkonium
spectrum at zero temperature.
The temperature dependence of the string tension was taken from
\cite{kaczmarek00}. The resulting quarkonia masses are shown in Fig \ref{th}.

Since the smallest distance available
on lattice is only $0.25 T^{-1}$ one may worry about possible medium effects at this
distance and their role in determination of $V_{\infty}(T)$. 
Normalizing the Polyakov loop correlator (\ref{polcorr}) at $r=1/(4T)$ with
Eq.\ (\ref{vTdep}), we thus obtain what might be a more reliable estimate
of the plateau $V_{\infty}(T)$ than with the $T=0$ form (\ref{cornell}).
It turns out, however,
that the two forms of
short distance behavior resulting from the zero temperature Cornell
potential  (\ref{cornell}) and (\ref{vTdep})
are practically identical, so that the normalization is in fact not affected
by the in-medium modifications at larger distances.
To consider further possible uncertainties of the normalization procedure,
we have also normalized the Polyakov loop correlator at the next smallest
distance $r=\sqrt{2}/(4T)$. The resulting two forms of $V_{\infty}(T)$
are shown in both Fig. \ref{th}. The difference
between the two curves of $V_{\infty}(T)$ provides an estimate of
the normalization error. Except for the region very near $T=T_c$,
the uncertainty is seen to be quite small.

From Fig. \ref{th} one can see that $\psi'$ and $\chi_c$ states 
become an open charm states well below $T_c$ and can dissociate
by decaying into $D \bar D$. The situation is similar for $\Upsilon(3S)$
and $\chi_b(2P)$ states which can decay into $B \bar B$
below $T_c$. For $J/\psi$, $\chi_b(1P)$ and $\Upsilon(2S)$
it is not possible to say whether they will dissociate above $T_c$ 
or just below $T_c$. Finally, the $\Upsilon(1S)$ state will definitely
dissociate above the deconfinement.

\section{Quarkonium dissociation by color screening}
\begin{figure}
\vspace*{-0.5cm}
\epsfxsize=8cm
\centerline{\epsffile{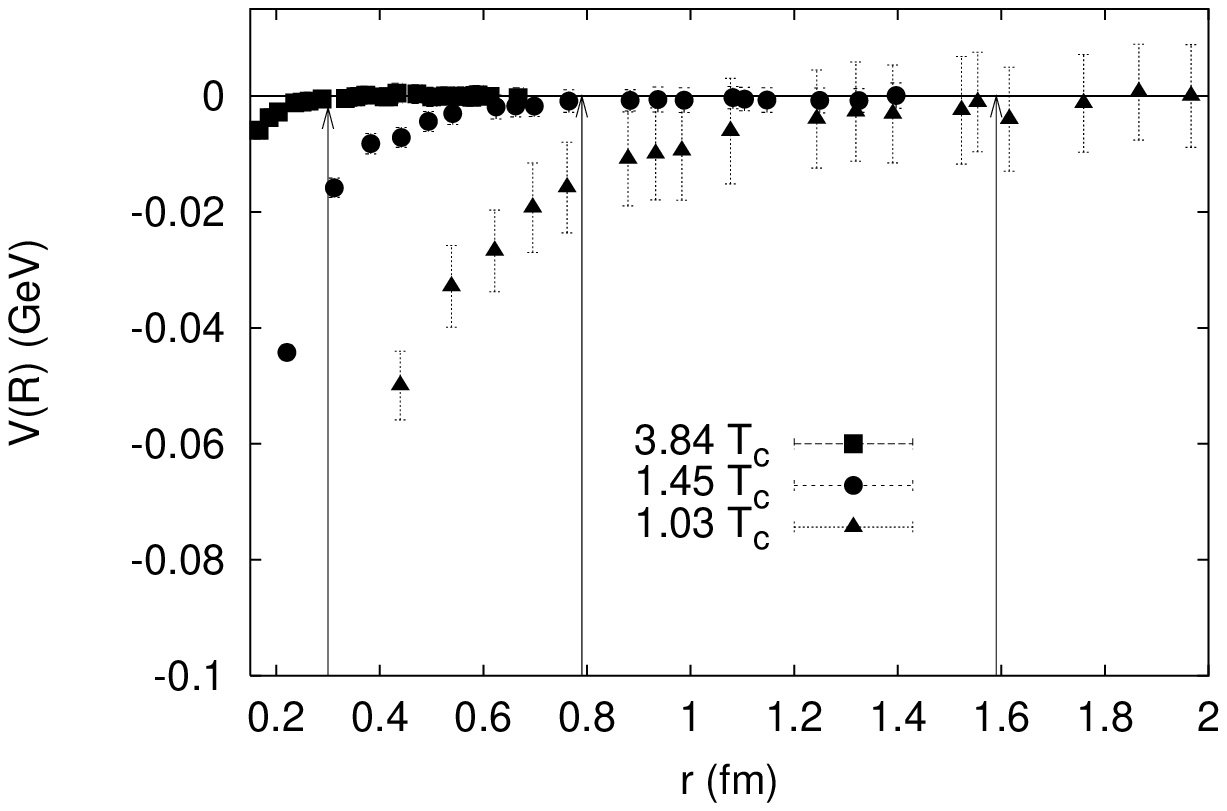} \hspace*{0.4cm} \epsfxsize=8cm \epsffile{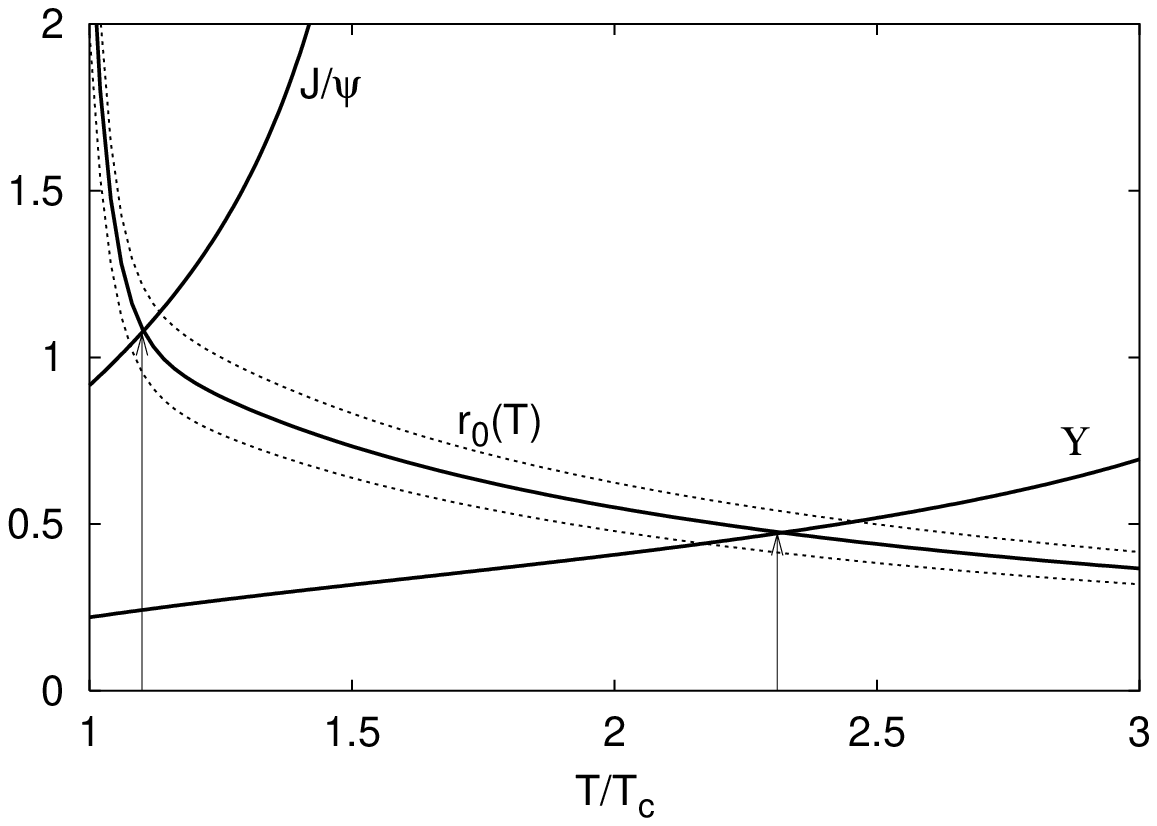}}
\vspace*{-0.3cm}
\caption{The color average potential calculated on lattice 
(left) and effective radii of the $J/\psi$ and $\Upsilon$ states (right)}
\label{above}
\end{figure}

In the deconfined phase it is customary to choose the constant $C$
in (\ref{polcorr}) to be the value of the correlator at infinite separation
$C=T \ln<L(r) L^{\dagger}(0)> \equiv {|<L>|}^2$.
The resulting connected correlator defines the so-called color
averaged potential \cite{nadkarni86}
\be
V(r,T)=-T \ln \frac{<L(r) L^{\dagger}(0)>}{{|<L>|}^2}
\label{vav}
\ee
The color average potential can be written as the thermal 
average of the potentials in color singlet $V_1(T,r)$ 
and color octet $V_8(T,r)$ states:
\be
\exp(-V(r,T)/T)=\frac{1}{9}\exp(-V_1(r,T)/T)+\frac{8}{9}\exp(-V_8(r,T)/T)
\label{av}
\ee
In potential models it is assumed that quarkonium is dominantly a singlet 
$Q\bar Q$ state. Furthermore the octet channel is repulsive (at least in
perturbation theory) and therefore only a singlet $Q\bar Q$  pair can be bound
in the deconfined phase. Thus we need to know the singlet potential.
The lattice data in the relevant case of 3 flavor QCD exist only for the
averaged potential \cite{karsch01,private}. The averaged potential in 3 flavor QCD 
are shown in Fig. \ref{above} for three representative temperatures.
Note that within present accuracy of the lattice calculations the potential
vanishes beyond some distance $r_0(T)$ denoted by vertical  arrows in Fig. \ref{above}.
In perturbation theory, the leading terms
for both are at high temperature and small $r$ ($r << T^{-1}$) of
Coulombic form,
\be
V_1(T,r) = -{4\over 3} {\alpha(T) \over r}, ~~~
V_8(T,r) = +{1\over 6} {\alpha(T) \over r},
\label{3.9}
\ee
with $\alpha(T)$ for the temperature-dependent running coupling.
In the region just above the deconfinement point $T=T_c$, there will
certainly be significant non-perturbative effects of unknown form.
We therefore first consider the high temperature regime, which we
somewhat arbitrarily define as $T \ge 1.45~T_c$. In this region, we attempt
to parameterize the existing non-perturbative effects through a
conventional screening form, replacing Eq.\ (\ref{3.9}) by
\be
-{3\over 4}V_1(T,r) = 6~V_8(T,r) = {\alpha(T) \over r} \exp\{-\mu(T)r\},
\label{3.10}
\ee
where $\mu(T)$ denotes the effective screening mass in the deconfined
medium.
We fit the lattice data by (\ref{av}/\ref{3.10}) assuming
$\alpha(T)$ and $\mu(T)$ to be unknown functions of $T$.
Such a fit can describe the lattice data for $T>1.45T_c$ very well. Furthermore the temperature dependence of $\alpha(T)$ can be well described by 1-loop
formula for the coupling in QCD with $\Lambda_{QCD}=(0.34 \pm 0.01)T_c$ and
the screening mass $\mu(T)$ turns out to be constant in units of the temperature
$\mu(T) = (1.15 \pm 0.02)  T$. A similar behavior of the screening mass was found in pure SU(2)
and SU(3) gauge theory \cite{heller95} - \cite{cucchieri01}.
We shall now assume that the above form of the screening mass
continues to remain valid as we lower the temperature to $T_c$.
Such a
constant screening mass down to $T_c$ is again expected from studies of
pure gauge theory \cite{heller95}.
On the other hand, quenched QCD (pure SU(3) gauge theory) studies
indicate that when $T$ is lowered to $T_c$, the perturbative ratio
$V_1/V_8 = - 8$ will increase in favor of the singlet potential
\cite{attig88}. We therefore try to describe the behavior just above
$T_c$ by a potential of the form (\ref{av}), in which the color octet
potential is given by
\be
V_8(T,r) = {c(T)\over 6} {\alpha(T) \over r} \exp\{-\mu r\}
\label{3.12}
\ee
instead of Eq.\ (\ref{3.10}); the factor $c(T) \leq 1$ accounts for the
expected reduction of octet interactions as $T \to T_c$. In the interval
$T_c < T <1.45~T_c$ we thus fit the lattice results for $V(T,r)$ in terms of
the parameter $c(T)$ with $\alpha(T)$ and  $\mu(T)$ given above.
In this way we get again quite a good fit of the lattice data  
Refs. \cite{karsch01,private} at temperatures $T<1.45 T_c$.

Now we are in a position to discuss quarkonium dissociation due
to color screening. It is natural to assume that the heavy $Q \bar Q$
pair cannot exist as a bound state if its effective binding radius (
the mean distance between $Q$ and $\bar Q$) is larger than the screening
radius of the medium. The effective radii for different bound states are
calculated from the Schr\"odinger equation with the singlet potential 
described above.
\be
\left[ 2m_a + {1\over m_a}\nabla^2 + V_1(r) \right] \Phi_i^a = M_i^a
\Phi_i^a,
\label{4.1a}
\ee
The screening radius of the medium can be identified with $1/\mu(T)$.
However, the value of $\mu(T)$ strongly depends on assumption we have
made to determine it. A more model independent and more conservative
approach would be to identify the screening radius with $r_0(T)$
defined above. We use the latter approach. 
In Fig. \ref{above} we show the effective radius of $J/\psi$ and 
$\Upsilon$ states and the screening radius as function of the 
temperature. The intersection of these curves defines the dissociation
temperature of $J/\psi$ and $\Upsilon$ states. Similar analysis was done
for excited states which may survive above $T_c$.

\section{Summary and Conclusions}
\begin{figure}
\begin{center}
a \hspace*{7.4cm} b\\
\vspace*{-0.6cm}
\includegraphics[width=18pc]{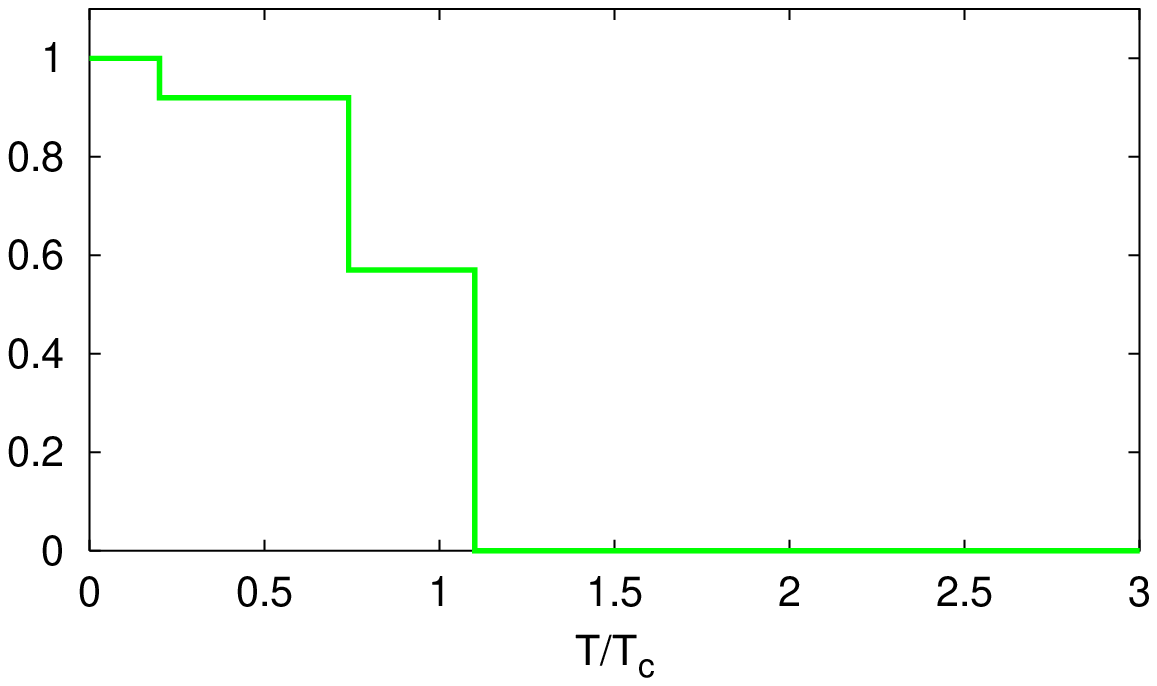}
\includegraphics[width=15pc]{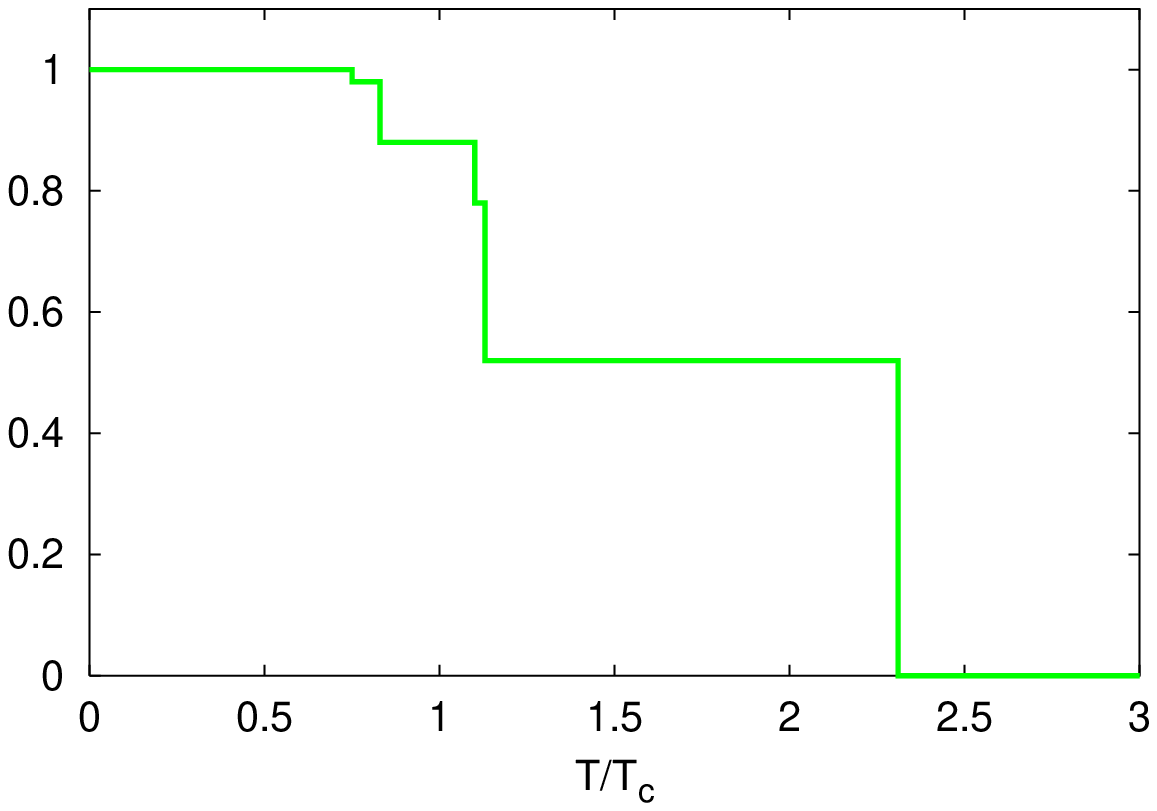}
\end{center}
\caption{
The suppression pattern of $J/\psi$ (left) and $\Upsilon$ yield (right)
as function of the temperature.
}
\label{supp}
\end{figure}

We have considered quarkonium dissociation in hot strongly 
interacting matter below as well as above the deconfinement. In a confined
medium dissociation of certain quarkonium states occur due to in-medium 
modification of the open charm (beauty) threshold as well as the quarkonia masses.
In the deconfined medium quarkonium dissociation is due to color screening. We summarize
the dissociation temperature of different quarkonium states in Tab. 2. Combining
these dissociation temperature with the feed-down fractions determined in section II
we can predict the sequential suppression pattern of $J/\psi$ and $\Upsilon$ states
as function of the temperature.
These are summarized in Fig. \ref{supp}.
For a more accurate determination of the quarkonium suppression
patterns, it would be desirable to carry out direct lattice studies of
the color singlet potential and of its quark mass dependence, which
may become important near the critical temperature. Furthermore, to make
contact with nuclear collision experiments, a more precise determination
of the energy density via lattice simulations is clearly needed, as is
a clarification of the role of a finite baryochemical potential.
For the latter problem, lattice studies are so far very difficult;
nevertheless, a recent new approach \cite{fodor} could make such
studies feasible.

The methods used in the present study of in-medium modification of
hadrons containing heavy quarks is of limited validity. A more accurate
study of hadron properties in medium should be based on determination
of hadron spectral function and their temperature dependence.
Recent lattice studies indicate that a determination of hadron spectral 
function is possible at least within quenched approximation with
the present days computer resources \cite{mem}. Although at present the hadronic
spectral function can be determined only in quenched approximation they
may provide valuable test for the approach described here. In principle such studies
will  also
include the dissociation of quarkonium states due to interaction with partonic
constituents (thermal activation) in terms of width of the peaks in the calculated
spectral function which is not accessible within present approach.
\vspace*{0.2cm}
\begin{center}
\begin{tabular}{|c|c|c|c|c|c|c|c|c|}
\hline
$q\bar{q}$&$J/\Psi$&$\chi_c $&$\psi'$&$\Upsilon(1S)$&$\chi_b
(1P)$&$\Upsilon (2S)$&$\chi_b (2P)$&$\Upsilon (3S)$ \\
\hline
$T/T_c$&1.10 & 0.74 & 0.2& 2.31& 1.13& 1.10& 0.83& 0.75\\
\hline
\end{tabular}
\vspace*{0.2cm}

TAB 2: Dissociation temperatures of different quarkonium states.
\end{center}

\noindent {\bf Acknowledgments:}
\noindent
The work has been supported by the 
DFG under grant FOR 339/1-2 and by BMFB under grant 06 BI 902.


\begin{thebibliography}{30}
\bibitem{matsui86}
T. Matsui and H. Satz, Phys. Lett. {\bf B178} (1986) 416
\bibitem{digal01a}
S. Digal et al, Phys. Lett. {\bf B514} (2001) 57
\bibitem{karsch01}
F. Karsch et al, Nucl. Phys. {\bf B605} (2001) 579
\bibitem{digal01b}
S. Digal et al, Phys. Rev. {\bf D64} (2001) 094015
\bibitem{cobb78}
J.\ H. Cobb et al., \PL {\bf 72} (1978) 497;\par
C. Koukoumelis et al., \PL {\bf B81} (1979) 405
\bibitem{lem82} 
Y.\ Lemoigne et al., \PL {\bf B113}  (1982) 509
\bibitem{ant92}
L.\ Antoniazzi et al., \PR {\bf D46} (1992) 4828;\par
\PRL {\bf 70} (1993) 383
\bibitem{cdf}
F. Abe et al.\ (CDF), \PRL {\bf 75} (1995) 4358; 
T.\ Affolder et al.\ (CDF), \PRL {\bf 84} (2000) 2094; CDF
Note 5027
\bibitem{mangano}
See e.g.\ M.\ Mangano, hep-ph/9507353
\bibitem{gavai95}
R.\ Gavai et al., Int.\ J.\ Mod.\ Phys.\ {\bf A10} (1995)
3043
\bibitem{bali00}
G.S. Bali, Phys. Rept. {\bf 343} (2001) 1
\bibitem{bali97}
G.S. Bali et al, Phys. Rev. {\bf D56} (1997) 2566
\bibitem{eichten80}
E.\ Eichten et al., \PR {\bf D17} (1978) 3090; \PR {\bf D21}
(1980) 203.
\bibitem{gao}
M.\ Gao, \PR {\bf D40} (1989) 2708.
\bibitem{kaczmarek00}
O.\ Kaczmarek et al., \PR {\bf D62} (2000) 034021
\bibitem{nadkarni86}
S. Nadkarni, \PR {\bf D33} (1986) 3738
\bibitem{private}
F.\ Karsch and E.\ Laermann, private communication

\bibitem{heller95}
U.\ M.\ Heller et al, Phys. Lett. {\bf B355} (1995) 511.

\bibitem{heller98}
U.\ M.\ Heller et la, Phys. Rev. {\bf D57} (1998) 1438.

\bibitem{cucchieri01}
A.\ Cucchieri et al., Phys. Rev. {\bf D64} (2001) 036001

\bibitem{attig88}
N.\ Attig et al, Phys. Lett. {\bf B209} (1988) 65

\bibitem{fodor}
Z. Fodor and S.D. Katz, hep-lat/0104001

\bibitem{mem}
I. Wetzorke et al, hep-lat/0110132
\end{thebibliography}
\end{document}